\documentclass[%
superscriptaddress,
reprint,
amsmath,amssymb,
aps,
prl,
10pt,
twocolumn,
floats,
]{revtex4-2}

\usepackage[english]{babel}
\usepackage[utf8]{inputenc}
\usepackage[T1]{fontenc}
\usepackage{graphicx}
\usepackage[caption=false]{subfig}
\usepackage{float}
\usepackage{dcolumn}
\usepackage{blindtext}
\usepackage{upgreek}
\usepackage{siunitx}
\usepackage{xcolor}

\DeclareSIUnit\gauss{G}
\newcommand{\ket}[1]{\left|#1\right\rangle}
\newcommand{\kcirc}[1]{\ket{#1 \mathrm{C}}}

\newcommand{\mean}[1]{\left\langle #1 \right\rangle}

\begin{document}

\title{Array of Individual Circular Rydberg Atoms Trapped in Optical Tweezers}

\author{B.~Ravon}
\altaffiliation{These authors contributed equally to this work.}
\author{P.~M\'ehaignerie}
\altaffiliation{These authors contributed equally to this work.}
\author{Y.~Machu}
\author{A.~Durán~Hernández}
\author{M.~Favier}
\author{J.~M.~Raimond}
\author{M.~Brune} 
\author{C.~Sayrin}
\email[Corresponding author: ]{clement.sayrin@lkb.ens.fr}

\affiliation{Laboratoire Kastler Brossel, Coll\`ege de France, CNRS, ENS-Universit\'e PSL, Sorbonne Universit\'e, 11 place Marcelin Berthelot, F-75231 Paris, France}

\date{\today}

    \begin{abstract}
Circular Rydberg atoms (CRAs), i.e., Rydberg atoms with maximal orbital momentum, are highly promising for quantum computation, simulation and sensing. They combine long natural lifetimes with strong inter-atomic interactions and coupling to electromagnetic fields. Trapping individual CRAs is essential to harness these unique features. We report the first demonstration of CRAs laser-trapping in a programmable array of optical bottle beams. We observe the decay of a trapped Rubidium circular level over $\SI{5}{\milli\second}$ using a novel optical detection method. This first optical detection of alkali CRAs is both spatially- and level selective We finally observe the mechanical oscillations of the CRAs in the traps. This work opens the route to the use of circular levels in quantum devices. It is also promising for quantum simulation and information processing using the full extent of Rydberg manifolds.
    \end{abstract}

\maketitle    
 
Quantum technologies exploit individual quantum systems to overcome the limitations of classical devices. Quantum cryptography, sensing, communication and computing rely on the manipulation of individual photons, ions, molecules, artificial or neutral atoms... Among these systems, Rydberg atoms~\cite{Gallagher1994, Adams2019} are well suited for quantum computing and simulation~\cite{Saffman2010, Saffman2016}, whether digital~\cite{Isenhower2010, Wilk2010} or analog~\cite{Browaeys2020}, quantum optics~~\cite{Firstenberg2016, Peyronel2012, Baur2014, Gorniaczyk2014, Distante2016, Busche2017, Ripka2018, Li2019} and sensing~\cite{Sedlacek2012a, Facon2016, Kumar2017, Cox2018}. With a high-lying valence electron (principal quantum number $n \gg 1$), these atoms are strongly coupled to electromagnetic fields and have huge dipole-dipole interactions, with  $\simeq$~10--100 $\si{\mega\hertz}$  interaction frequencies for atoms micrometers apart. 
 
Many experimental platforms operate with arrays of Rydberg atoms~\cite{Gaetan2009, Ebert2014, Labuhn2016, Endres2016, Zeiher2017,
 Kim2020, Madjarov2020}. Quantum simulations with a few hundred interacting Rydberg atom, in low-angular-momentum ($\ell$) levels, have,e.g., recently been performed, with atoms initially trapped in optical tweezers (OTs)~\cite{Scholl2021, Ebadi2021}. The simulation time is, however, limited to a few microseconds. The natural lifetime of low-$\ell$ levels is a few $\SI{100}{\micro\second}$ only. In a microsecond, one atom out of a hundred is lost and relaxation hampers the simulator operation. 

Circular Rydberg atoms (CRAs), with a maximal orbital momentum (magnetic quantum number $|m| = \ell = n-1$), have much longer lifetimes, about $\SI{30}{\milli\second}$ for $n=50$, hundred times those of low-$\ell$ levels with the same $n$. Recent proposals suggest using regular arrays of these long-lived CRAs for quantum simulation and quantum computing~\cite{Nguyen2018, Meinert2020a, Cohen2021}. They promise, in particular, simulation of spin dynamics over unprecedented timescales. To benefit from the long CRA lifetimes, one must trap them. The interaction-induced motion of free atoms would, otherwise, change the interatomic distances and the interaction frequencies within microseconds. 

Laser trapping of  alkali atoms Rydberg levels relies on the positive ponderomotive energy~\cite{Dutta2000} that they experience in a laser field. Recently, experiments have demonstrated the ponderomotive trapping of low-$\ell$ Rydberg levels in 3D~\cite{Graham2019, Barredo2020} and of circular Rydberg levels in 2D~\cite{Cortinas2020}. Low-$\ell$ Rydberg levels of alkaline earth atoms have also been trapped by using dipole forces on the ionic core~\cite{Wilson2022}. However no experiment so far has combined the long-lived CRAs with long-term, ordered optical traps.

Here, we demonstrate the preparation of a regular array of individual laser-trapped Rubidium circular Rydberg levels. We observe trapping over a few millisecond timescale. First, ground-state atoms are trapped in an array of optical tweezers, i.e., tightly-focused Gaussian beams acting as attractive dipole traps. We excite these atoms to a circular  level and transfer them into an array of repulsive ponderomotive traps, made up of hollow optical bottle beams (BOBs), with an intensity minimum surrounded by a light shell~\cite{Barredo2020}. We develop a CRA optical detection method, both spatially- and level-selective. We use it to observe microwave-induced Rabi flopping between two circular levels and the motion of the CRAs in the traps. This work is a decisive step in the development of quantum technologies using CRAs.

 \begin{figure*}[t]
 \centering
 \includegraphics[width=.9\linewidth]{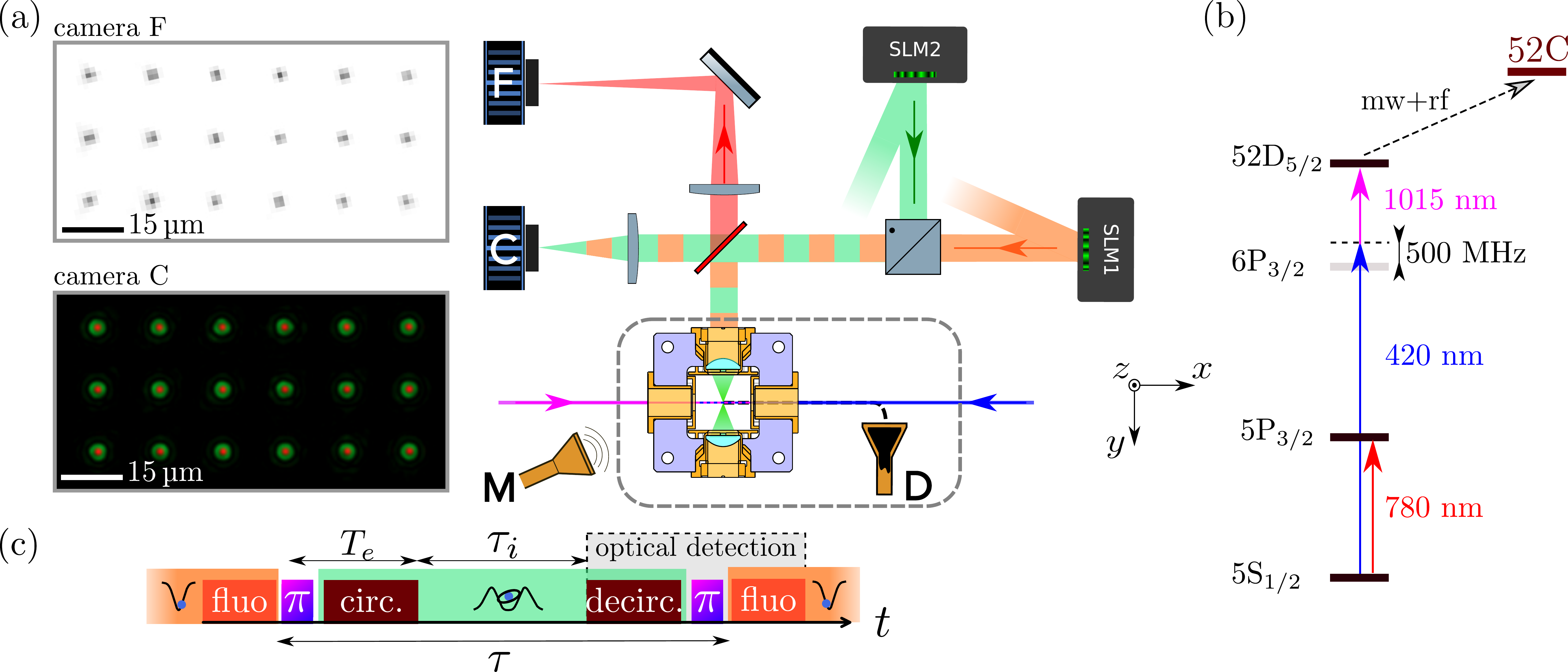}
 \caption{(a) \textbf{Experimental setup} A sapphire cube (violet) holds electric-field control electrodes (orange) and two short-focal-length lenses (cyan). One of them focuses two beams at $\SI{820}{\nano\meter}$ (green), the phase profiles of which are tailored by two SLMs. The same lens collects $780$-$\si{\nano\meter}$ light (red) emitted by the trapped atoms trapped. The Rydberg excitation lasers (magenta and blue) enter the cube along the $x$ axis. A microwave horn (M) shines MW field into the UHV chamber (grey dashed square). A channeltron (D) counts Rb ions guided towards it (black dashed line) by electrostatic lenses (not shown). Two cameras collect the fluorescence photons (F) and control the intensity profiles of the trapping beams. Insets: (F) fluorescence image of the 18 trapping sites averaged over $1000$ sequences; (C) The OT (red) and BOB (green) beam overlap is controlled on the camera. (b) \textbf{Level diagram} of $^{87}\mathrm{Rb}$ displaying the fluorescence line and the Rydberg excitation transitions. (c) \textbf{Typical experimental sequence} Atoms are loaded in the OTs (orange) and imaged by fluorescence (red). A two-photon $\pi$ pulse excites them to Rydberg states. They are then trapped in BOBs (green) and  transferred to circular Rydberg states (brown). The procedure is reversed to image the atoms at the end of the sequence. The optical detection block (grey) may be replaced by field-ionization detection. The time $\tau$ is the time during which the OTs are turned off during an experimental run, and the idle time $\tau_i$ is the time between circular state preparation and detection.}
 \label{fig:setup}
\end{figure*}

The experimental setup is depicted in Fig.~\ref{fig:setup}(a). Rubidium-87 atoms are loaded from a 2D-magneto-optical trap (MOT) into a 6-beam 3D-MOT at the center of a hollow sapphire cube, held in an UHV chamber. The cube holds electrodes, which control the electric field in 3D, and two short-focal-length ($f=\SI{16.3}{\milli\meter}$) aspheric lenses, which focus a $\SI{820}{\nano\meter}$-wavelength laser beam at the center of the cube. A liquid-crystal spatial light modulator~\cite{SLM} [SLM1, Fig.~\ref{fig:setup}(a)] imprints on this beam the phase mask required to obtain a $3\times 6$ regular array of OTs, separated by $\SI{15}{\micro\meter}$. It also corrects the optical aberrations of the setup. The $1.2$-$\si{\micro\meter}$-waist and $2.6$-$\si{\milli\watt}$-power  red-detuned OTs form 1-$\si{\milli\kelvin}$-deep dipole traps. They are turned on during the operation of the 3D-MOT and the subsequent optical molasses stage. Individual Rb atoms are loaded in the collisional-blockade regime~\cite{Schlosser2002} into the OTs, with an average filling of $0.62$.

To detect the atoms, we use fluorescence imaging on the $\ket{5S_{1/2}, F=2} \to \ket{5P_{3/2}, F'=3}$ transition [level diagram in Fig.~\ref{fig:setup}(b)] using the MOT beams tuned $\SI{60}{\mega\hertz}$ below resonance for an atom at the center of the OT. Scattered photons are collected by the focusing lens and recorded by an EMCCD camera over $\SI{15}{\milli\second}$. A single-shot image allows us to discriminate empty sites from those containing one atom. This fluorescence imaging heats the atoms and is thus followed by an optical molasses stage. 

Prior to the excitation to Rydberg levels, the trapped atoms are optically pumped to $\ket{5S_{1/2}, F=2, m_F=2}$ using a $\sigma^+$-polarized laser field on resonance with the $\ket{5S_{1/2}, F=2} \to \ket{5P_{3/2}, F'=3}$ transition. The quantization axis is set by a $B=\SI{7}{\gauss}$ magnetic field aligned along the $x$-axis [see Fig.~\ref{fig:setup}(b)]. The temperature of the atoms after this stage is $\SI{23}{\micro\kelvin}$~\cite{Suppl}. For some experiments, we further cool the atoms down to $\SI{7}{\micro\kelvin}$ by adiabatically lowering the trap depth by a factor of 10 within $\SI{500}{\micro\second}$. We restore it just before fluorescence imaging.

\begin{figure*}[t]
\centering
	 \includegraphics[width=.90\linewidth]{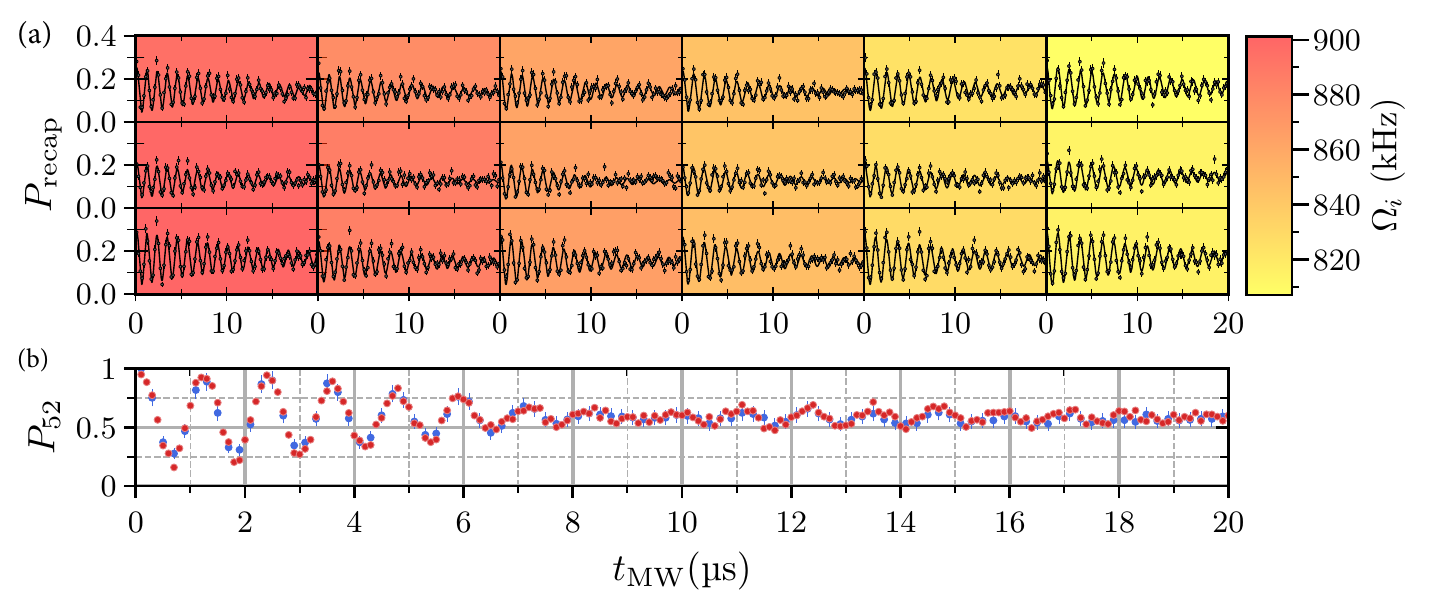}
\caption{\textbf{Rabi oscillations.} A MW pulse (duration $t_\mathrm{MW}$) drives the $\kcirc{52}\to\kcirc{50}$ transition. (a) Recapture probability (black points) $P_\mathrm{recap}$ measured for the 18 sites with the optical detection method. The background color indicates the value of the Rabi frequency $\Omega_i$ obtained from a fit of the data to a damped sine. (b) Probability $P_\mathrm{52}$ (see text) to find an atom in $\kcirc{52}$ averaged over the array, measured with field ionization (red points) or optical detection (blue points).}
\label{fig:Rabi} 
\end{figure*}

We excite the atoms to the $\ket{52D_{5/2}, m_J = 5/2}$ Rydberg state in a 2-photon process, with two laser beams at $420$-$\si{\nano\meter}$ and $1015$-$\si{\nano\meter}$. Both are $\sigma^+$ polarized and detuned by $\SI{500}{\mega\hertz}$ from $\ket{6P_{3/2}, F'=3, m_F=3}$.  The OTs are turned off before this excitation  to avoid inhomogeneous light shifts of the transition. Once the atoms are in the Rydberg level, we turn on an array of $3\times6$ BOBs, superimposed with the OT array via a polarizing beamsplitter ahead of the focusing lens. The BOBs originate from the same laser source as the OTs and are tailored by a second SLM [SLM2 in Fig.~\ref{fig:setup}(a)]~\cite{Barredo2020, Suppl}. SLM2 is also used to align in 3D the BOBs with the OTs, ensuring an efficient recapture of the atoms. The $\SI{20}{\milli\watt}$ power per BOB results in a trap depth for Rydberg states $\approx \SI{70}{\micro\kelvin}$.

We transfer the atoms from $\ket{52D_{5/2}, m_J=5/2}$ to $\ket{52F, m = 2}$ by a resonant $1.9$-$\si{\micro\second}$-long microwave (MW) $\pi$-pulse at $\SI{64.8}{\giga\hertz}$, shone into the setup via the MW horn M. We then set an electric field $\mathbf{F}$, aligned with $\mathbf{B}$, to a value $F = \SI{2.4}{\volt\per\centi\meter}$. By ramping the electric field down to $\SI{2.1}{\volt\per\centi\meter}$ in the presence of a $\sigma^+$-polarized RF field at $\SI{225}{\mega\hertz}$, we adiabatically transfer~\cite{Cantat-Moltrecht2020} the atoms to the $n=52$ circular level, denoted $\kcirc{52}$. At room temperature, its lifetime is $\SI{130}{\micro\second}$, while the complete transfer from the ground state lasts $T_e = \SI{15}{\micro\second}$. The purity of the prepared $\kcirc{52}$ state is measured~\cite{Cantat-Moltrecht2020} to be about $0.9$. The preparation efficiency, from the ground state to $\kcirc{52}$, is estimated to be $0.70$, mainly limited by inhomogeneities of the laser, RF and MW excitation over the traps array.

Level-selective detection of CRAs can be performed by field-ionization~\cite{Cantat-Moltrecht2020}. We ionize the atoms in a  $\approx$~100~$\si{\volt\per\centi\meter}$ electric field and count the ions with the channeltron D, located outside the sapphire cube [Fig.~\ref{fig:setup}(a)]. This detection, however, is not spatially-selective. It cannot distinguish atoms in different BOBs. We thus develop a  spatially- and level selective optical detection. We transfer the atoms in $\kcirc{52}$ back to the $\ket{F=2, m_F=2}$ ground-state by time-reversing the $T_e$-long sequence used for the circular state excitation [Fig.\ref{fig:setup}(c)]. The BOBs are turned off right before the final 2-photon laser $\pi$ pulse and the OTs are turned back on right after it. The atoms initially in $\kcirc{52}$ are recaptured in the OTs. The atoms in any other Rydberg state are not addressed by the optical $\pi$ pulse, are not mapped to the ground state and are not re-captured by the OTs, which repel Rydberg states. The atoms not initially excited to Rydberg states move out of the OT trapping region under the action of the BOBs and are not recaptured either. Fluorescence imaging finally reveals the sites where an atom has been recaptured. The recapture probability, $P_\mathrm{recap}$, for each site measures the population $P_{52}$ in $\kcirc{52}$ at the beginning of the optical detection.

We assess the optical detection method by observing the Rabi flopping on the  $\kcirc{52}\to\kcirc{50}$ two-photon transition, driven by a  MW field  at $\SI{49.6}{\giga\hertz}$. We record $P_\mathrm{recap}$ as a function of the MW pulse duration, $t_\mathrm{MW}$, and plot it in Fig.~\ref{fig:Rabi}(a) for each of the 18 trapping sites. The time $\tau$ during which the OTs are turned off is fixed at $\SI{60}{\micro\second}$. The clear oscillations between the two circular levels shows that the optical detection acts as a spatial- and level-selective measurement. The maximum recapture probability, $\sim{0.35}$, is limited by the finite efficiencies of the optical, MW and RF pulses and by the finite lifetime of the Rydberg states.  

We also calculate the probability $\mean{P_\mathrm{recap}}$ averaged over the 18 sites and estimate the population $P_{52}(t_\textrm{MW})$ as $\mean{P_\mathrm{recap}}(t_\mathrm{MW})/\mean{P_\mathrm{recap}}(0)$. In Fig.~\ref{fig:Rabi}(b), we plot (blue points) $P_{52}(t_\textrm{MW})$ and compare it to the same probability measured by field ionization (red points). The very good agreement between the two data sets certifies the CRAs optical detection.

Fitting the individual oscillations of Fig.~\ref{fig:Rabi}(a) by an exponentially damped sine, we recover the individual Rabi frequencies, $\Omega_i$. We plot their values as a background colormap. The collapse and revival observed in the averaged signal of Fig.\ref{fig:Rabi}(b) result from the beating of the different Rabi frequencies measured here. The observed $\SI{1.18}{\mega\hertz\per\milli\meter}$ gradient indicates that the atomic array lies in the vicinity of a node of the uncontrolled MW-field mode structure defined by the surrounding electrodes. Our measurement reflects the spatial profile of the MW field.  The individual plots also reveal that the contrast of the Rabi oscillations damps quickly. The atoms in the center of the array, i.e., those that have most neighbors, are the most affected. We thus attribute this damping to interactions between the CRAs. 

\begin{figure}[t]
 \centering
 \includegraphics[width=.95\linewidth]{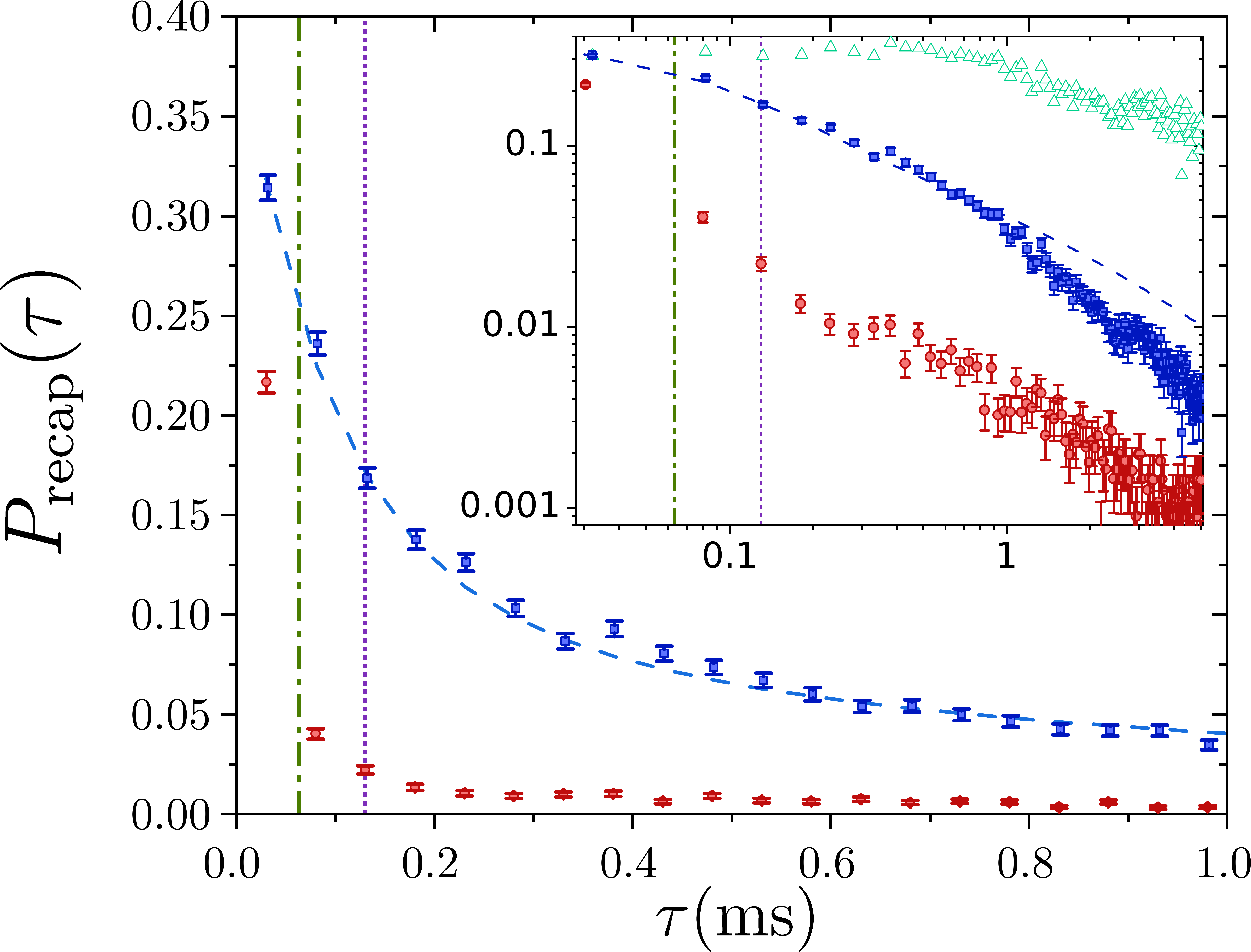}
 \caption{\textbf{Laser-trapping of CRAs} Recapture probability $P_\mathrm{recap}$ as a function of the time $\tau$ during which the OTs are turned off. Inset: log-log plot up to $\tau=\SI{5}{\milli\second}$. While the atoms are in the Rydberg states, the BOBs are either off (red circles) or on (blue squares). The calculated decay $P_\mathrm{lifetime}(\tau)$ is plotted as a blue dashed line, and the normalized probability $P_\mathrm{n}(\tau)$ as green triangles (inset). The lifetime of $\kcirc{52}$ and the oscillation period of the CRAs in the BOBs are indicated as dotted purple and green dashed-dotted vertical lines, respectively.}
 \label{fig:trapping}
\end{figure}

To demonstrate the optical trapping of CRAs, we prepare the atoms in $\kcirc{52}$ and optically detect them after a variable waiting time. We plot in Fig.~\ref{fig:trapping} the probability $P_\mathrm{recap}(\tau)$, averaged over the 18 sites, to recapture an atom as a function of the time $\tau$ during which the OTs are turned off. The atoms are left in $\kcirc{52}$ during the idle time $\tau_i = \tau - 2\times T_e$ that separates the end of circular state preparation and the start of the optical detection scheme [Fig.~\ref{fig:setup}(c)]. The minimal value of $\tau$ is $\tau_\mathrm{min} = \SI{32}{\micro\second}$, for which $\tau_{i, \mathrm{min}} = \SI{2}{\micro\second}$. The measurement is done either with (blue squares) or without (red circles) the BOB traps. Clearly, $P_\mathrm{recap}$ drops significantly slower with the BOBs on. Without them, the atoms fly away at their thermal velocity. The initial atomic temperatures are $\SI{7}{\micro\kelvin}$ or $\SI{23}{\micro\kelvin}$ for the blue and red data points, respectively. The atoms thus drift  by more than an OT waist within $\SI{19}{\micro\second}$ or $\SI{10}{\micro\second}$. If an atom in $\kcirc{52}$ leaves the trap or moves away from the OT location during $\tau_i$, it is de-excited back to its ground-state by the optical detection but not recaptured in the final OT. Applying the BOBs stops this expansion. The atoms are detected over a much longer time, proving that the CRAs have been trapped in the BOBs. 

We investigate the drop of $P_\mathrm{recap}$ within the BOBs by comparing it to the effect of the finite lifetime of $\kcirc{52}$, to which the level-selective optical detection is sensitive. We calculate the evolution, due to relaxation in free space at room temperature~\cite{Cantat-Moltrecht2020, Wu2023}, of the population $P_{52}(t)$ in $\kcirc{52}$, from an initially pure state. We plot (dashed line in  Fig.~\ref{fig:trapping}) the calculated values of $P_\mathrm{lifetime}(\tau) = P_{52}(\tau_i)\, P_\mathrm{lifetime}(\tau_\mathrm{min})$. It is in very good agreement with the measurements during the first millisecond. There is, thus, no visible loss of atoms from the BOBs during this time,  $8$ lifetimes of $\kcirc{52}$ and $16$ transverse oscillation periods of the atoms in the trap.

In the inset of Fig.~\ref{fig:trapping}, we plot in a log-log scale $P_\mathrm{recap}(\tau)$ at later times, up to $\tau=\SI{5}{\milli\second}$. After $\SI{1}{\milli\second}$, the measured damping of $P_\mathrm{recap}$ with the BOBs departs from the calculated decay. We also plot the normalized probability $P_\mathrm{n}(\tau) = P_\mathrm{recap}(\tau) / P_{52}(\tau_i)$ calculated with the BOBs only (green triangles). It remains constant over the first millisecond and then drops with a characteristic time $\approx\SI{5}{\milli\second}$, obtained by an exponential fit to the data. While this value is highly dependent on background-counts estimates~\cite{Suppl}, it can be considered as a coarse estimate of the trapping time of the CRAs in the BOBs.

\begin{figure}[t]
 \centering
 \includegraphics[width=.99\linewidth]{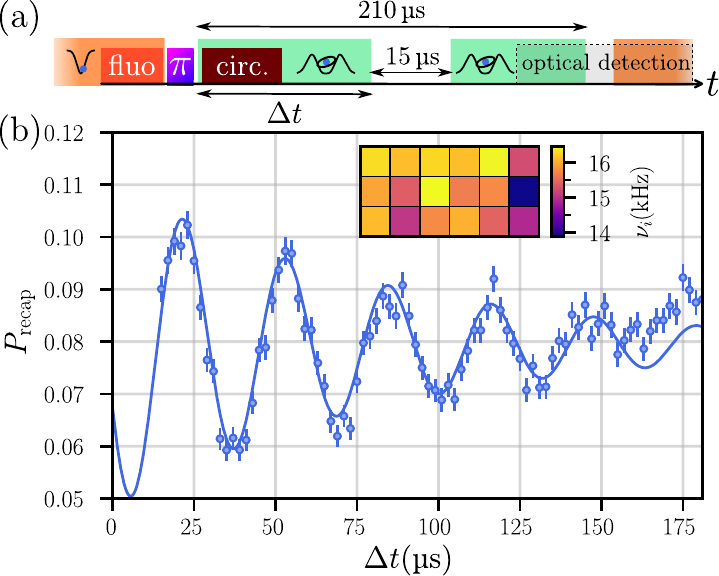}
 \caption{\textbf{Oscillations in the BOB traps.} (a) Experimental sequence: the BOB are turned off during $\SI{15}{\micro\second}$, at a time $\Delta t$ after the Rydberg excitation laser pulses. The CRAs are detected using the optical detection method. (b) Recapture probability $P_\mathrm{recap}$ averaged over the 18 sites (blue points). The solid line is a fit to a damped sine. The mosaic indicates the trapping frequency $\nu_i$ for each site, obtained from a fit of the individual signals. }
 \label{fig:bob_freq}
\end{figure}

We finally characterize the BOB traps by inducing oscillations of the CRAs within them. We displace the BOBs perpendicularly to the beam propagation direction by $\SI{300}{\nano\meter}$ with respect to the OT centers. The Rydberg atoms, initially at the edge of the BOBs, oscillate around the trap center. We turn off the BOBs during a fixed $\SI{15}{\micro\second}$ interval, starting $\Delta t$ after the BOBs have been turned on [Fig.~\ref{fig:bob_freq}(a)]. The time $\tau$ during which the OTs are off is set at $\SI{210}{\micro\second}$. We choose $\Delta t > T_e$ so that the atoms are in the circular states when the BOBs are switched off. 

Figure~\ref{fig:bob_freq} shows the measured recapture probability $P_\mathrm{recap}$ in the OT traps as a function of $\Delta t$, averaged over the 18 trapping sites. Oscillations of $P_\mathrm{recap}$ are clearly visible. If the atoms are at the BOB center, i.e., have maximal velocity at the turn-off time, they fly away from the trapping region and are not recaptured nor detected at the end of the sequence. If the atoms are at the edge of the trap at the turn-off time, they are recaptured and  detected. The probability $P_\mathrm{recap}(\Delta t)$ thus oscillates at twice the trapping frequency. The colormap in Fig.~\ref{fig:trapping} gives the estimated trapping frequencies $\nu_i$ for the 18 sites, obtained by a fit of the individual signals to a damped sine~\cite{Suppl}. The average oscillation frequency is $\SI{15.8}{\kilo\hertz}$, with a standard-deviation over the BOB array of $\SI{0.6}{\kilo\hertz}$ only. Using  a theoretical model of the BOB profile, this corresponds to a power of $\SI{19.9\pm2}{\milli\watt}$ per BOB, in fair agreement with our estimations~\cite{power}.

This work is the first demonstration of 3D optical trapping of individual circular Rydberg atoms in a regular array of traps. It enables the exploration of quantum information and quantum simulation protocols with interacting CRAs, that rely on long-term operation. 

We have developed an optical detection scheme of circular states both spatially- and level-selective. This technique will be decisive for the use of circular Rydberg states in quantum technologies. It is, in particular, required for the measurement of spin correlations over the array of traps, signatures of the quantum phases of an interacting-spin system~\cite{Nguyen2018}. It can also be used in devices that would combine the sensitivity of circular Rydberg atoms and the flexibility of optical imaging for, e.g., quantum sensing of magnetic or electric fields. We expect the efficiency of the method to be greatly improved by working in a cryogenic environment where circular states are long-lived~\cite{Cantat-Moltrecht2020}. Together with ponderomotive manipulation of the Rydberg states~\cite{Cardman2020, Malinovsky2020}, it opens he route towards a full-optical control of circular Rydberg atoms. 

Finally, our results are also relevant for experiments with other Rydberg levels than the circular ones. By allowing controlled interactions between Rydberg atoms over long times, it paves the way towards protocols, that exploit the richness of the Rydberg multiplicity~\cite{Kruckenhauser2022}.
 
\medskip
This work has been supported by the European Union FET-Flag project n\textdegree 817482 (PASQUANS), ERC Advanced grant n\textdegree\ 786919 (TRENSCRYBE) and QuantERA ERA-NET (ERYQSENS, ANR-18-QUAN-0009-04) and by the Région Île-de-France in the framework of DIM SIRTEQ (project CARAQUES).

\bibliography{bibliography}

\end{document}


\title{\textsc{Supplementary Material} \\ Array of Individual Circular Rydberg Atoms Trapped in Optical Tweezers}

\author{B.~Ravon}
\author{P.~M\'ehaignerie}
\author{Y.~Machu}
\author{A.~Durán~Hernández}
\author{M.~Favier}
\author{J.~M.~Raimond}
\author{M.~Brune} 
\author{C.~Sayrin}

\affiliation{Laboratoire Kastler Brossel, Coll\`ege de France, CNRS, ENS-Universit\'e PSL, Sorbonne Universit\'e, 11 place Marcelin Berthelot, F-75231 Paris, France}

\date{\today}

\maketitle    
 
\appendix

\section*{Phase mask}
\begin{figure}[h]
\begin{minipage}[t]{0.35\columnwidth}
\centering
\includegraphics[width=0.7\columnwidth]{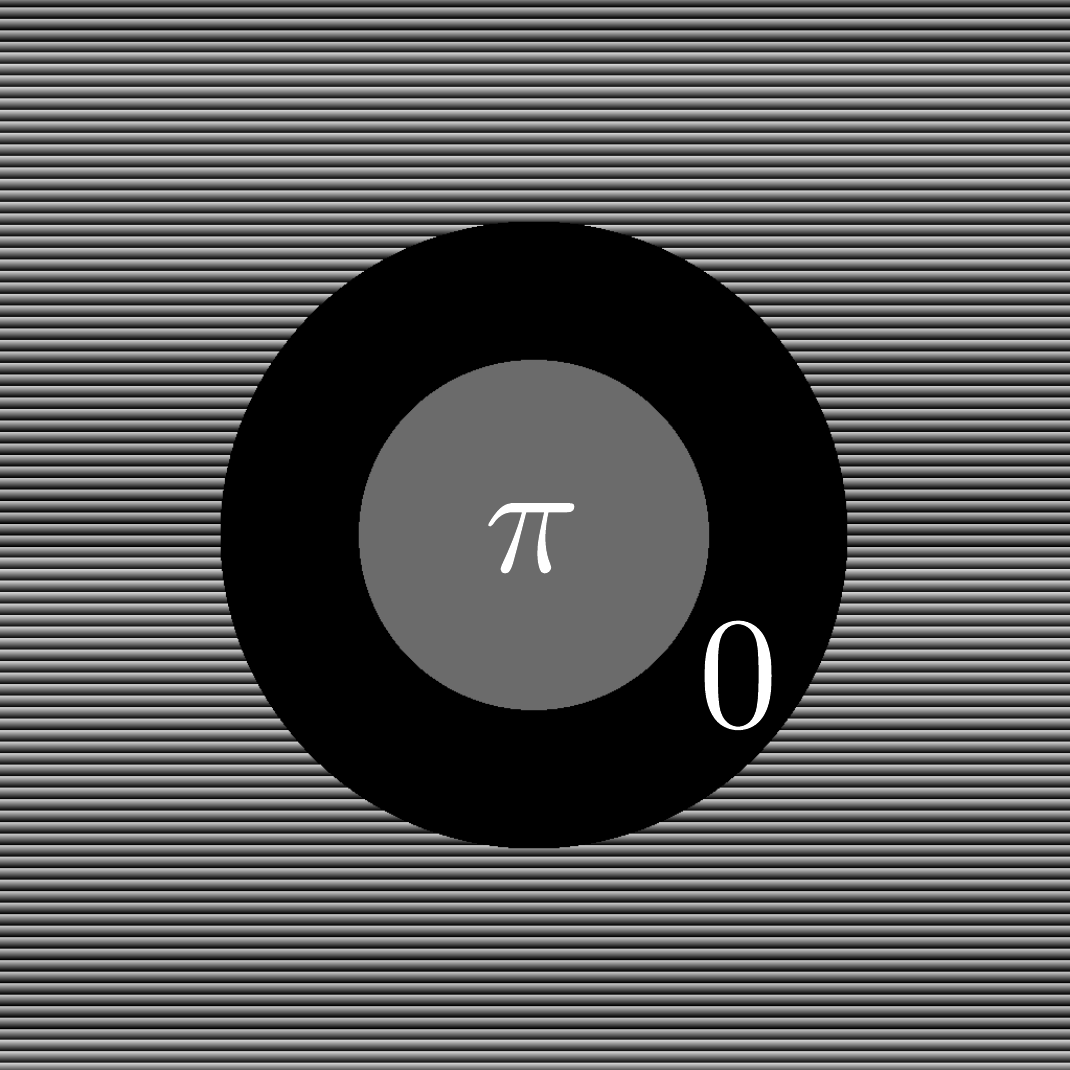}
\caption{Phase mask used to generate a single BOB. See text for details.}
\label{fig:suppl_phasemask}
\end{minipage}
\hfill
\begin{minipage}[t]{0.6\columnwidth}
\centering
\includegraphics[width=0.9\columnwidth]{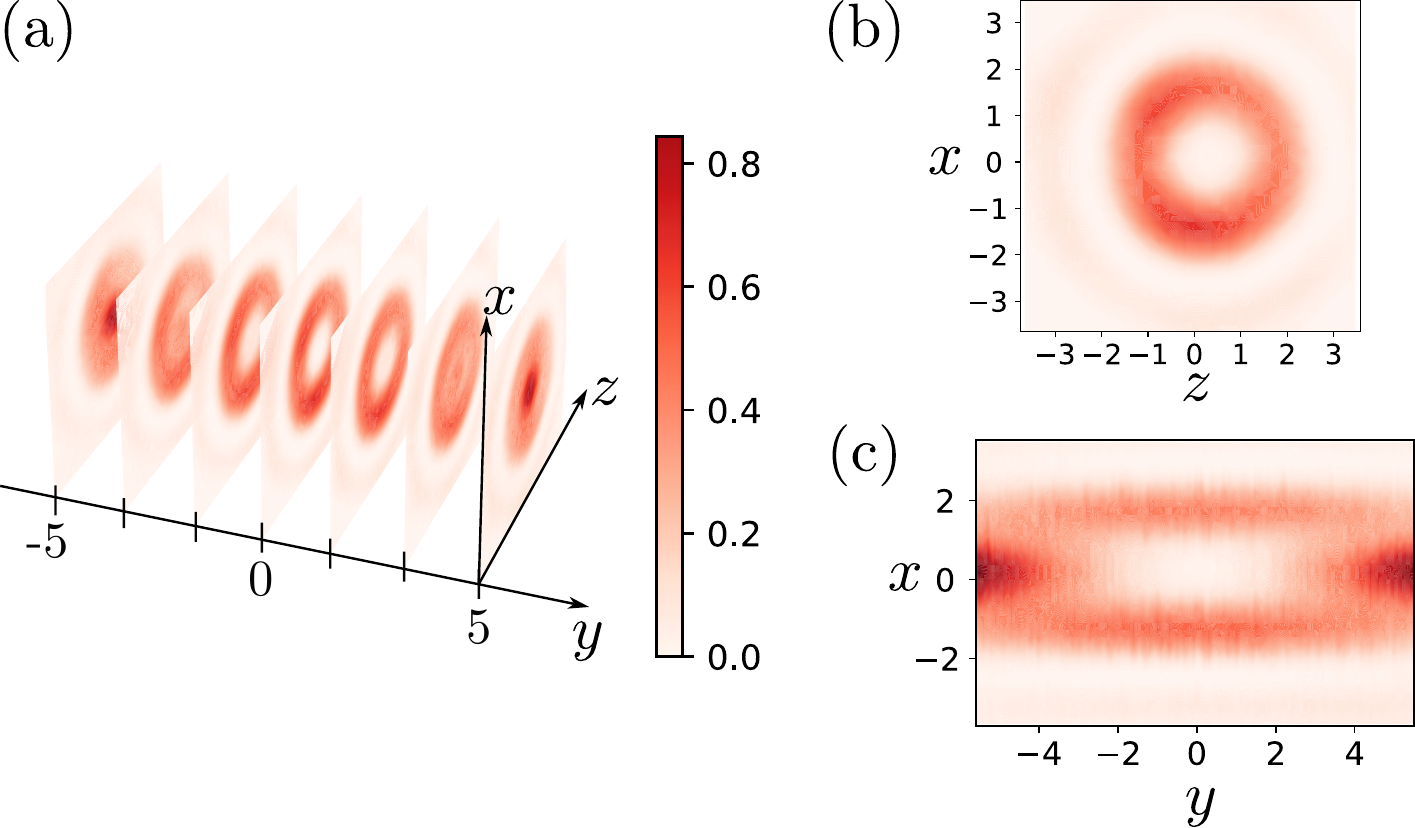}
\caption{Reconstruction of a BOB profile measured on a camera. All distances are given in microns. The colorscale indicates the BOB intensity as arbitrary units. It is shared by all three plots. (a) ($x$-$z$) cuts of the intensity profile measured in different $y$ positions along the propagation direction. (b) ($x$-$z$) cut measured at the center of the BOB. (c). ($x$-$y$) cut of the intensity profile. The plane contains the propagation axis.}
\label{fig:suppl_bobs}
\end{minipage}
\end{figure}

We show in Fig.~\ref{fig:suppl_phasemask} the ideal phase mask that is applied to SLM2 to engineer a single BOB. It consists of a $\pi$-phase disk surrounded by a $0$-phase disk. The ratio between the radii of the two disks is set to maximize the trapping depth. A linear gradient of phase is set out of the pupil defined by the larger disk. The incoming light that hits this zone is deflected out of the science region at the level of the atoms. To the ideal mask of Fig.~\ref{fig:suppl_phasemask}, we add inside the pupil a linear gradient of phase, orthogonal to the outside one, to separate the BOB diffracted in the first order from undiffracted light in the zero order. We also add the phase mask that turns a single BOB into an array of BOBs, and the ad hoc phase mask that corrects the optical aberrations of the optical setup.

We reconstruct the BOB intensity profile in 3D by measuring 2D cuts orthogonal to the propagation axis, i.e., the $y$-axis in Fig.~1 of the manuscript. These measurements are plotted in Fig.~\ref{fig:suppl_bobs}.

\section*{Measurement biases}
The renormalization process made in Figure~3 and the deduced trapping time of $\SI{5}{\milli\second}$ must be considered with caution. First, because the recapture probability drops to percents or less, the measurement gets sensitive to biases such as the recapture of ground-state atoms, initially in a different state than $\ket{F=2, m_F=2}$. These atoms are not sensitive to the Rydberg-excitation lasers, and may be trapped in the high-intensity region of the BOBs before being recaptured, though not efficiently, in the OTs. This systematic error significantly depends on the relative alignment between the BOBs and the OTs, and is estimated to be at most of a few $10^{-4}$. Second, the population $ P_{52}(t)$ is calculated considering a decay in free space. The surrounding electrodes, nevertheless, can modify the volume mode density in the relevant MW spectrum, as hinted by the MW field mapping of Fig.~2. A deeper analysis would require long-lived circular Rydberg levels obtained in a cryogenic environment. 

\section*{Atom temperature measurement}
\begin{figure}[t]
\centering
\includegraphics[width=0.45\columnwidth]{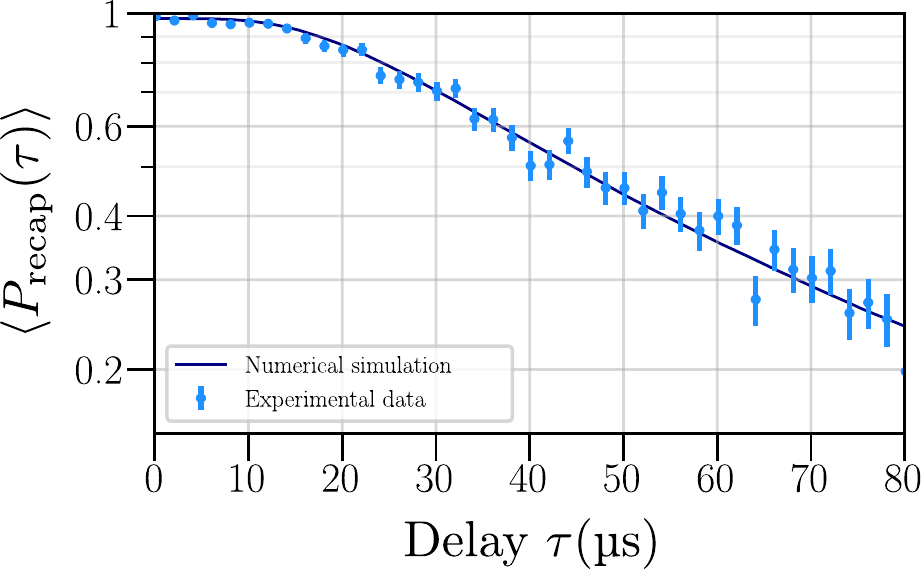}
\caption{Measurement of the temperature of ground state atoms. The recapture probability is measured as a function of the turn-off time $\tau$. The solid line is a fit to the data (see text).}
\label{fig:suppl_temp}
\end{figure}
The measurement of the temperature of individual atoms in the OT traps uses a standard release-recapture technique. We turn off the OTs during a variable time $\tau$ and measure the subsequent probability to recapture the atom in the OT traps. The data is then fitted to a Monte-Carlo simulation with the temperature $T$ as the only free parameter. It sets the initial position and velocity distribution of the atoms in the trap. We plot in Fig.~\ref{fig:suppl_temp} the outcome of such a measurement made after the adiabatic cooling of the atoms (reduction by a factor 10 of the OT power). The fit to the data corresponds to a temperature of $T= \SI{6.6\pm1}{\micro\kelvin}$. Note that the measurement is done for the atoms in their ground-state. The Rydberg excitation itself kicks the atoms by about $\SI{6}{\milli\meter\per\second}$, namely $\approx 0.23$ times the thermal velocity for atoms at $\SI{7}{\micro\kelvin}$.

\section*{Oscillations in BOBs}
\begin{figure*}[b]
\centering
\includegraphics[width=0.9\columnwidth]{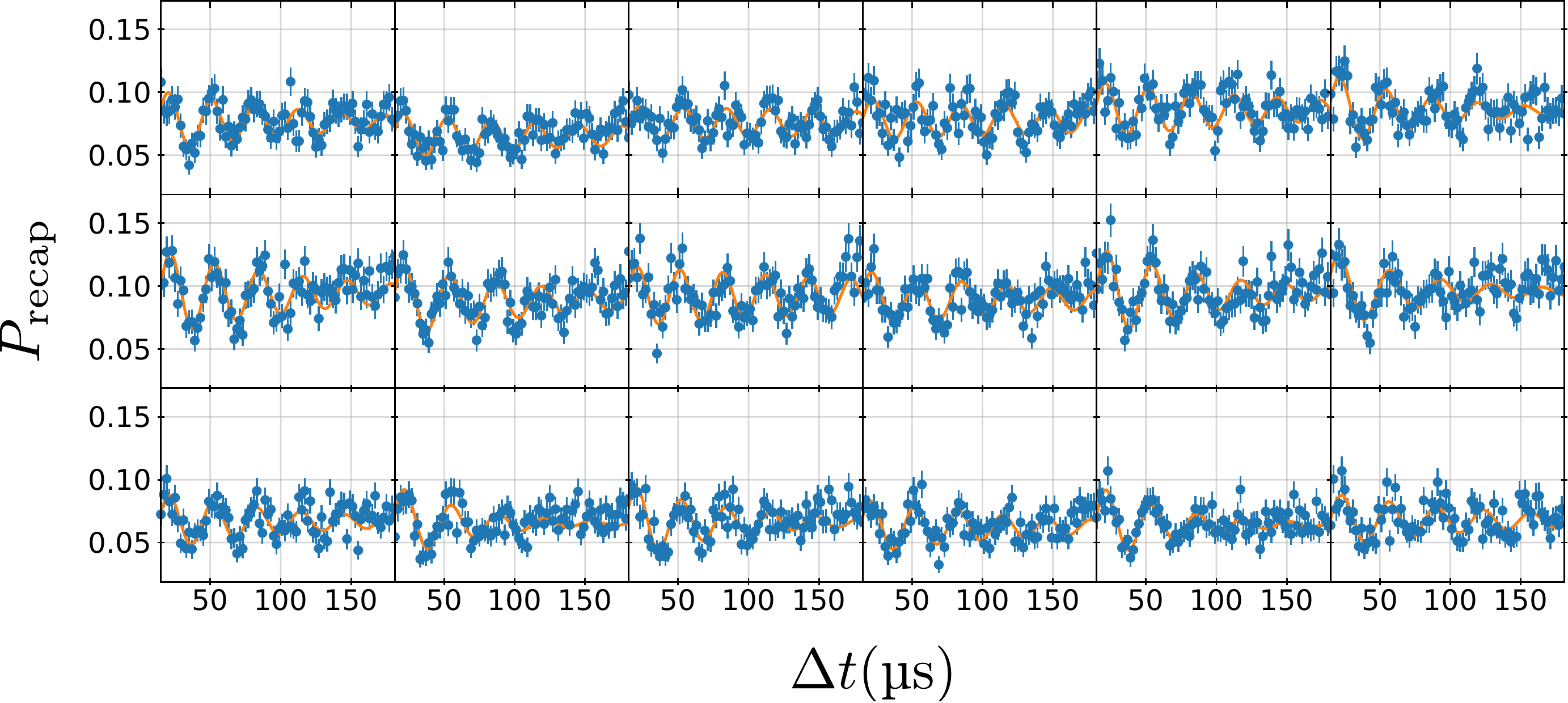}
\caption{Individual signals corresponding to the measurement of the oscillation frequency of the atoms in the BOB traps, see Fig~4 of the manuscript. The recapture probability is measured as a function of the delay time $\Delta t$ (see manuscript).}
\label{fig:suppl_osc}
\end{figure*}
We plot in Fig.~\ref{fig:suppl_osc} the individual signals that correspond to the measurement of Figure~4. The orange lines correspond to fits to the data with a damped sine. The red line in Fig.~4 of the manuscript is the average of these fits. The obtained oscillation frequencies are given in the mosaic of Fig.~4 of the manuscript.